\pdfoutput=1

\documentclass{emulateapj}
\usepackage{amsmath,amssymb,graphicx}

\newcommand{\Ob}{\Omega_b}
\newcommand{\Om}{\Omega_m}
\newcommand{\Ol}{\Omega_\Lambda}
\newcommand{\Mpch}{{\rm Mpc}\ h^{-1}}
\newcommand{\kpch}{{\rm kpc}\ h^{-1}}
\newcommand{\Msunh}{M_\odot\ h^{-1}}
\newcommand{\zreion}{z_{\rm reion}}
\newcommand{\lya}{{\rm Ly}\alpha}

\newcommand{\bx}{{\bf x}}

\shorttitle{Imprint of Inhomogeneous Hydrogen Reionization}
\shortauthors{Trac, Cen, \& Loeb}

\begin{document}

\title{Imprint of Inhomogeneous Hydrogen Reionization on the\\ Temperature Distribution of the Intergalactic Medium}

\author{Hy Trac$^1$, Renyue Cen$^2$, \& Abraham Loeb$^1$} 

\affil{$^1$ {\it Harvard-Smithsonian Center for Astrophysics, Cambridge, MA
02138, USA}\\ $^2$ {\it Department of Astrophysical Sciences, Princeton
University, Princeton, NJ 08544, USA}}

\begin{abstract}

We study the impact of inhomogeneous hydrogen reionization on the thermal evolution of the intergalactic medium (IGM) using hydrodynamic + radiative transfer simulations where reionization is completed either early ($z\sim9$) or late ($z\sim6$). In general, we find that low-density gas near large-scale overdensities is ionized and heated earlier than gas in the large-scale, underdense voids. Furthermore, at a later time the IGM temperature is inversely related to the reionization redshift because gas that is heated earlier has more time to cool through adiabatic expansion and Compton scattering. Thus, at the end of reionization the median temperature-density relation is an inverted power-law with slope $\gamma-1\sim-0.2$, in both models. However, at fixed density, there is up to order unity scatter in the temperature due to the distribution of reionization redshifts. Because of the complex equation-of-state, the evolved IGM temperature-density relations for the redshift range $4\lesssim z\lesssim6$ can still have significant curvature and scatter. These features must be taken into account when interpreting the Ly$\alpha$ absorption in high redshift quasar spectra.

\end{abstract}


\section{Introduction}

The cosmic reionization of hydrogen fundamentally changes the thermal and ionization conditions for the high redshift ($z\gtrsim6$) intergalactic medium (IGM), converting a cold and neutral gas into a warm and highly ionized one \citep[see][for a review]{2008Loeb}. During this inhomogeneous process, the radiative transfer (RT) of the UV field, primarily generated by stellar sources, will proceed such that large-scale, overdense regions near sources are generally ionized and heated earlier than large-scale, underdense regions far from sources \citep[e.g.][]{2004BarkanaLoeb, 2004WyitheLoeb, 2004FZH, 2008Lee}. The photo-ionization of hydrogen (HI $\rightarrow$ HII) and helium (HeI $\rightarrow$ HeII) is expected to heat the IGM to temperatures of $\gtrsim10^4$ K, with higher temperatures obtained for harder radiation spectra \citep[e.g.][]{1990MiraldaEscude, 1994MiraldaEscude}. Thereafter, the thermal evolution will be governed by adiabatic cooling/heating for underdense/overdense gas, Compton scattering on the Cosmic Microwave Background (CMB), atomic line cooling, shock-heating, and additional photo-heating only where recombination is efficient.

In principle, the temperature of the IGM can be quantified through the thermal broadening of the Lyman alpha ($\lya$) forest lines \citep[e.g.][]{1999Schaye, 2000Schaye, 2000Ricotti, 2001McDonald} and the Jeans smoothing of the $\lya$ forest flux power spectrum \citep[e.g.][]{2000Theuns, 2001Zaldarriaga, 2004Viel}. Measurements of the temperature can then be used to put constraints on when the reionization of hydrogen and helium occurred \citep[e.g.][]{2002Theuns, 2003HuiHaiman}. Note that the full reionization of helium (HeII $\rightarrow$ HeIII) is expected to occur at
lower redshifts ($z\sim3$), triggered by quasars with harder spectra \citep[see][for recent modelling]{2008Furlanetto,
2008bBolton, 2008McQuinn}. The current observational constraints have large uncertainties, although better measurements will come in time and improved models will be required to interpret them.

In this {\it Letter} we focus on the impact of inhomogeneous hydrogen reionization on the thermal evolution of the IGM. In particular, we study the equation-of-state and measure the temperature-density relation and its scatter using hydrodynamic + RT simulations. \citet{1997HuiGnedin} have studied the effects of uniform reionization and found that the temperature-density relation is well approximated by a positive power-law slope for the low-density IGM. However, inhomogeneous reionization can give rise to a much different equation-of-state, especially right after completion. Specifically, the large-scale, underdense voids are initially hotter than higher density gas near sources, that having been ionized and heated earlier have also had more time to cool. This gives rise to an equation-of-state characterized by an inverted (negative slope) temperature-density relation. At later times, the underdense gas cools adiabatically faster and the temperature contrast decreases and eventually reverses back to the positive slope.

\section{Simulations}

\begin{figure*}[t]
\center
\includegraphics[width=3.4in]{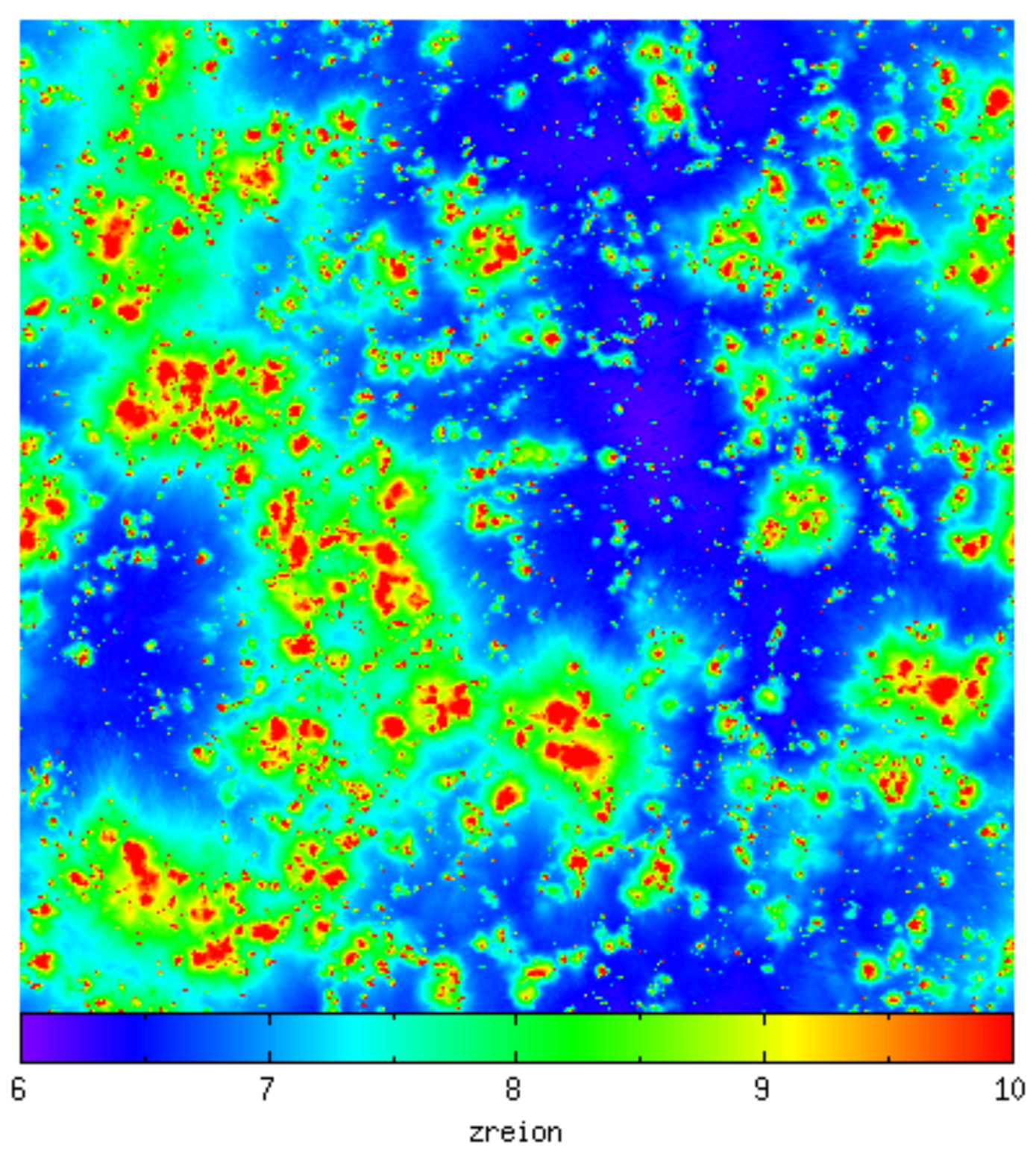}\includegraphics[width=3.4in]{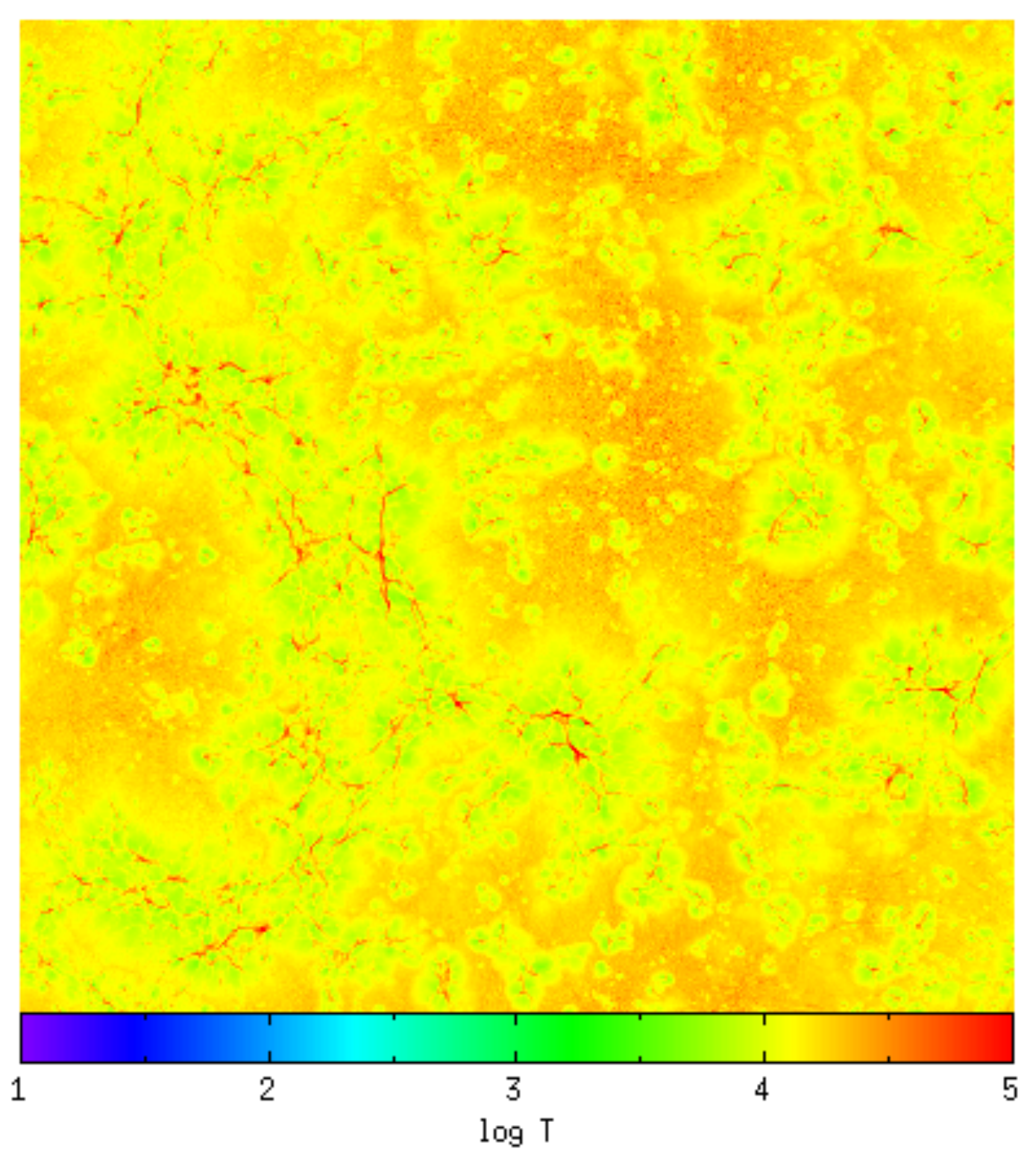}
\caption{Low-resolution images of a slice of size $(100\ \Mpch)^2$ with a thickness of two hydro cells (130 $\kpch$) from the late reionization model. {\it Left}: Redshift at which Jeans-size elements of the IGM get reionized. {\it Right}: Temperature field at the end of reionization. In general, low-density gas that is ionized and heated earlier has lower temperature at late times.}
\label{fig:maps}
\end{figure*}

We adopt a hybrid approach in modelling cosmic reionization that allows us to cover a larger dynamic range and satisfy the requirements of having sufficiently high resolution to capture small-scale structure, and a sufficiently large volume to reduce sample variance \citep{2004BarkanaLoeb}. Below, we summarize two recent numerical simulations incorporating N-body, hydrodynamic, and RT algorithms to solve the coupled evolution of the dark matter, baryons, and radiation \citep{2004TracPen, 2006TracPen, 2007TracCen}. The simulations are based on the recommended WMAP 5-year cosmological parameters: $\Om=0.28$, $\Ol=0.72$, $\Ob=0.046$, $h=0.70$, $\sigma_8=0.82$, and $n_s=0.96$ \citep{2008Dunkley}.

Our hybrid approach is divided into two major components. The first task involved running a high-resolution N-body simulation with $3072^3$ dark matter particles on an effective mesh with $11520^3$ cells in a comoving
box, $100\ \Mpch$ on a side. We identified collapsed dark matter halos on the fly using a friends-of-friends algorithm, with a linking length $b=0.2$ times the mean interparticle spacing, in order to model radiation sources and sinks. With a particle mass resolution of $2.68\times10^6\ \Msunh$, we can reliably locate all dark matter halos with virial temperatures above the atomic cooling limit ($T\sim10^4$ K) with a minimum of $\sim40$ particles \citep{2007Heitmann}, and half of this collapsed mass budget is resolved with $>400$ particles per halo. Our halo mass functions are in very good agreement with other recent work \citep[e.g.][]{2007ReedBFJT, 2007LukicHHBR, 2008CohnWhite}.

The second component of the work produces a series of hydrodynamic + RT simulations with moderate resolution, but incorporating subgrid physics modelled using the high-resolution information from the large N-body simulation. For the first two simulations in the series, we consider basic models where reionization is completed early ($z\sim9$) and late ($z\sim6$) to roughly reflect our current state of knowledge on the details of the reionization process \citep[e.g.][]{2003cenb, 2003WyitheLoeb, 2003HaimanHolder, 2006FCK}. Radiation sources are prescribed and star formation rates calculated using the halo model described in \citet{2007TracCen}. Here we consider only Population II stars from starbursts \citep{2003Schaerer} as contributing to the ionizing photon budget. We also neglect subgrid clumping and self-shielding of dense absorbers such as minihalos or Lyman limit systems, but will study their effects in ongoing work.

Each simulation utilizes equal numbers ($N=1536^3$) of dark matter particles, gas cells, and adaptive rays, where for the latter, we track 5 frequencies above the hydrogen ionizing threshold of 13.6 eV. The photo-ionization and photo-heating rates for each cell are calculated from the incident radiation flux and used in the non-equilibrium solvers for the ionization and energy equations. The initial conditions are generated with a common white noise field and a linear transfer function calculated with CAMB \citep{2000Lewis}.  The simulations were run using the NASA Columbia Supercomputer.

\section{Results}

\begin{figure*}[t]
\center
\includegraphics[width=3.3in]{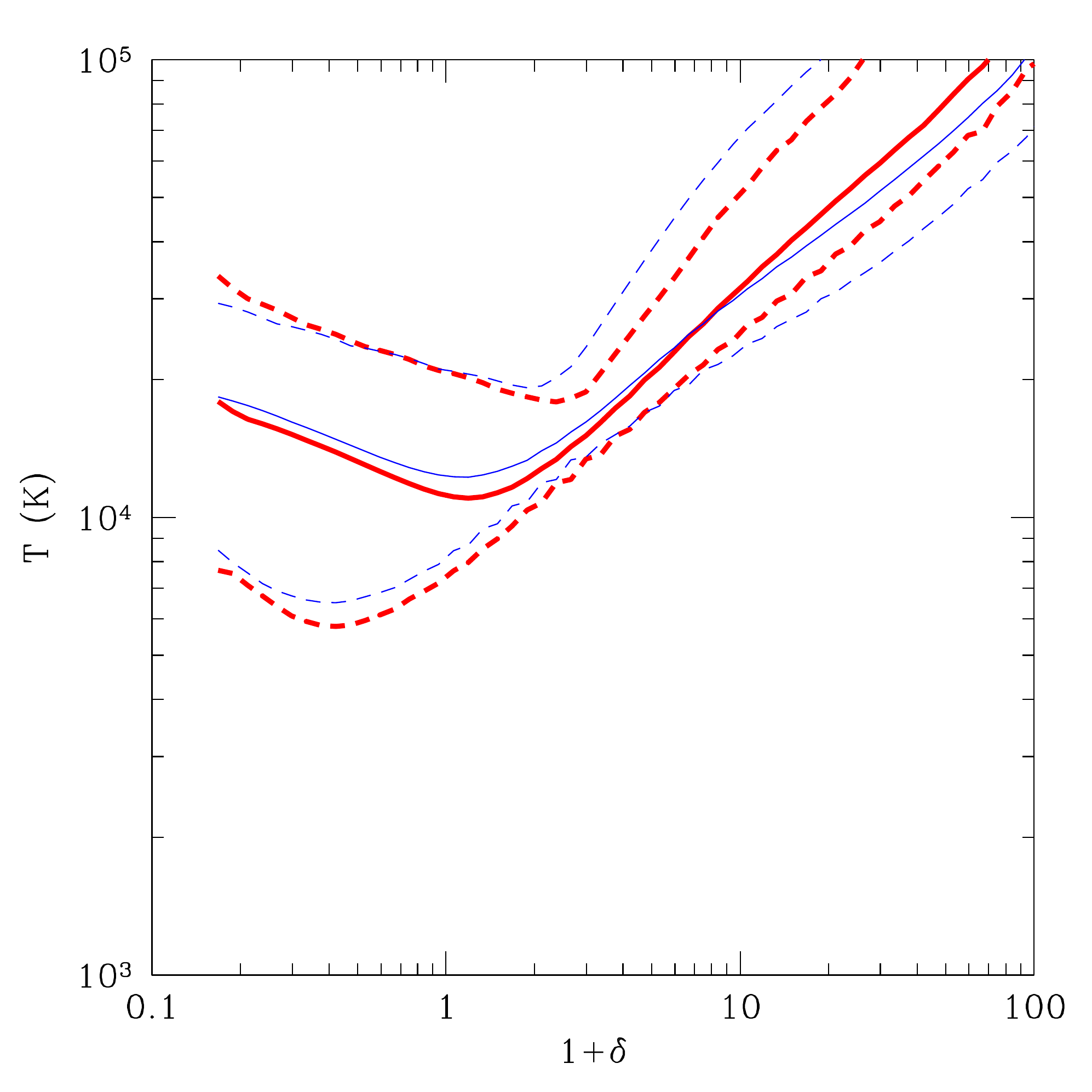}\includegraphics[width=3.3in]{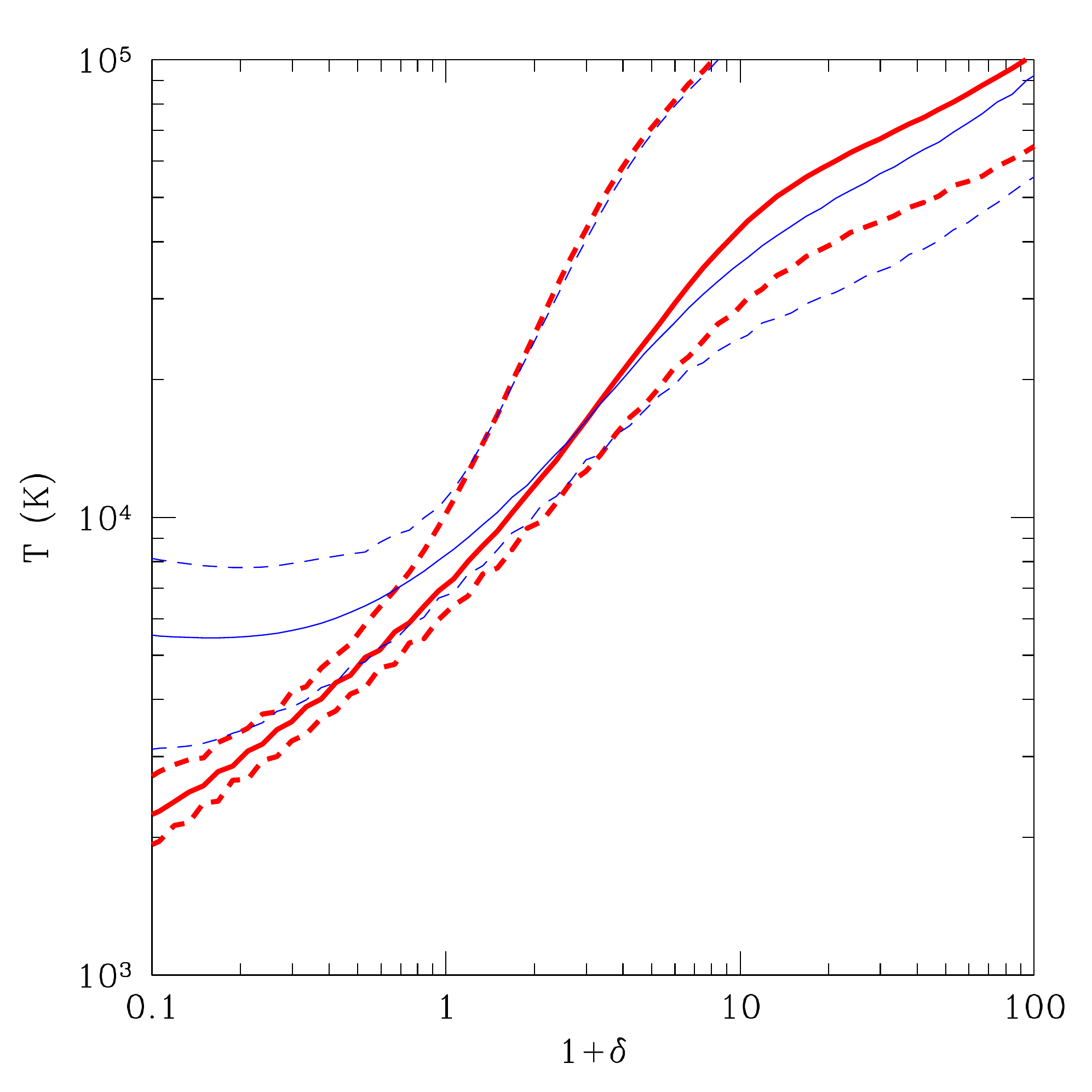}
\caption{Gas temperature-density relations at the end of reionization ({\it left}) and at a redshift $z=4$ ({\it right}) for the early ({\it red, thick lines}) and late ({\it blue, thin lines}) reionization models. For each density bin of width 0.05 dex, we plot the median temperature ({\it solid lines}) and the 2-$\sigma$ spread ({\it dashed lines}). At the end of reionization, the two models have similar inverted power-law $T$-$\rho$ relations for the low-density IGM, with slopes $\gamma-1\sim-0.2$ and order unity scatter. By $z=4$, the early model has settled into a rising power-law with only small scatter. However, the late model is closer to yielding an isothermal equation-of-state with still significant scatter of thousands of kelvins.}
\label{fig:T_of_d}
\end{figure*}

\subsection{The Redshift of Reionization}

Over the course of each simulation we track when a gas cell first becomes more than 99\% ionized, enabling us to construct a 3D reionization-redshift field $\zreion(\bx)$ in parallel to  the gas density $\rho(\bx)$ and temperature $T(\bx)$ fields. In Fig.~\ref{fig:maps} we show a slice from the late reionization simulation box to illustrate the inhomogeneity of the reionization process and the large-scale correlations. The fields have been smoothed on cells of comoving length $130\ \kpch$, which is close to the Jean's or filtering scale \citep{1998GnedinHui, 2000bGnedin}. In general, $\zreion$ is highly correlated with the large-scale density field, as HII regions originate around biased, overdense sources and final overlap occurring in the large-scale, underdense voids.  We find that the strong and positive correlation extends down to scales $\sim 1\ \Mpch$.

The temperature field at the end of reionization shows considerable complexity with some interesting characteristics. First, we note the presence of cool ($T\lesssim10^4$ K), low-density gas just outside the shock-heated filamentary and halo gas. Second, the low-density gas gets progressively warmer ($T\gtrsim10^4$ K) as one moves away from sources and towards the voids. It is apparent from the maps that the temperature is inversely related to $\zreion$ for the low-density IGM and we will quantify this relation in the next section.

\subsection{The Temperature-Density Relation}

\begin{figure}[b]
\center
\includegraphics[width=3.3in]{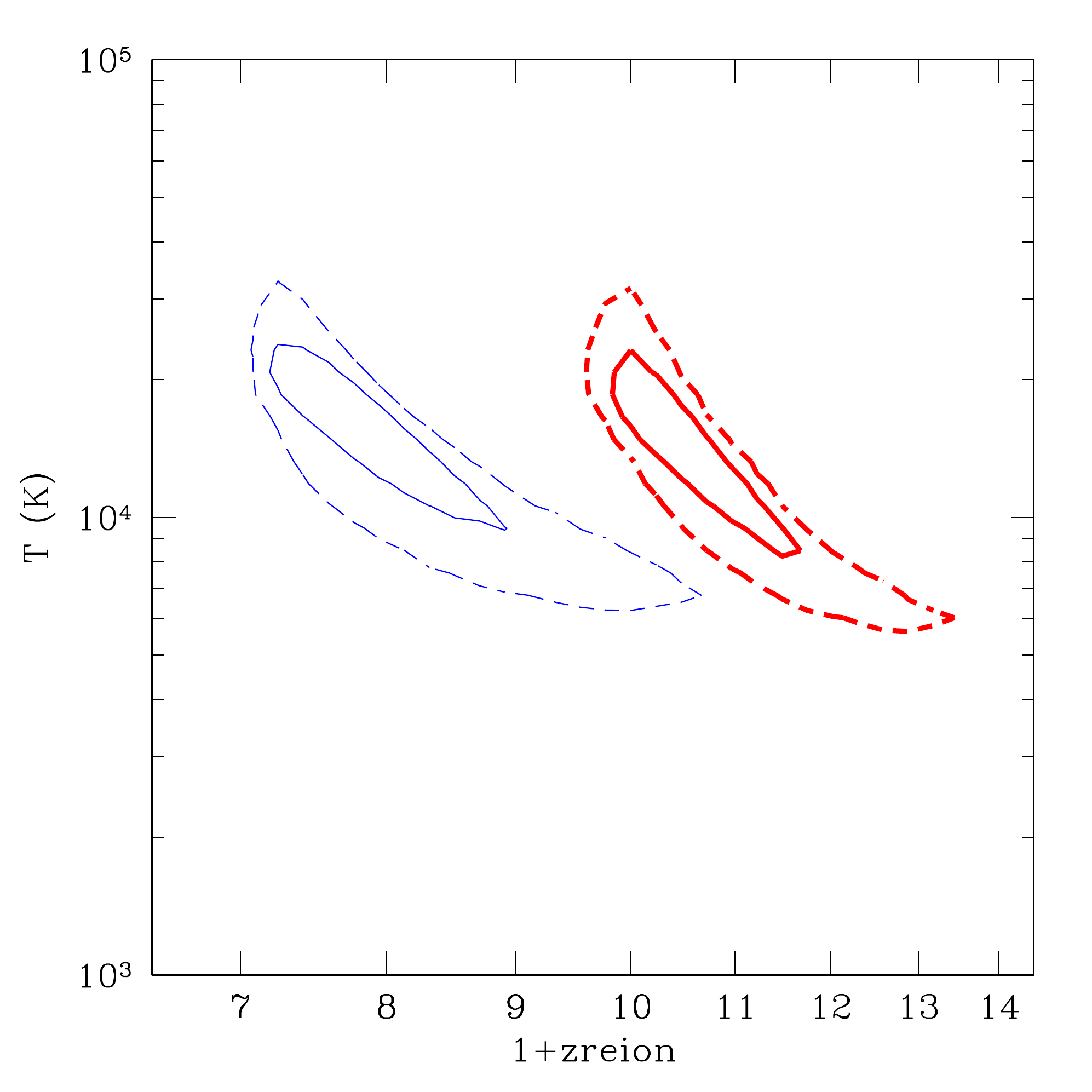}
\caption{End-of-reionization temperature as a function of the reionization redshift for low-density ($\delta=-0.5\pm0.05$) gas, for the early ({\it red, thick lines}) and late ({\it blue, thin lines}) models. The two contours specify the 1-$\sigma$ ({\it solid lines}) and 2-$\sigma$ ({\it dashed lines}) boundaries for the cumulative probability distributions.}
\label{fig:T_of_z}
\end{figure}

Fig.~\ref{fig:T_of_d} shows the gas temperature-density relations at the end of reionization and at redshift $z=4$ for the early and late reionization models. For each density bin of width 0.05 dex, we plot the median temperature and the 2-$\sigma$ spread around it. At the end of reionization, the two models have complex but similar $T$-$\rho$ relations,
driven by photo-heating, shock-heating, and cooling. For the low-density IGM where photo-heating dominates over shock-heating, we find that the median temperature approximately follows a power-law relation,
\begin{equation}
T = T_0(1+\delta)^{\gamma-1},
\end{equation}
with a slope $\gamma-1\sim-0.2$. The median temperature has an inverted form, but the difference in the median temperatures at $\delta=0$ and $\delta=-0.8$ is small compared to the scatter in the temperature for underdense gas. The order unity scatter in the temperature at a fixed density is explained below.

We first note that the IGM density on the Jean's scale is only weakly correlated with the density averaged on large scales, which are still in the linear regime. low-density gas is also found close to large-scale overdense regions around sources and it preferentially gets ionized and heated earlier than gas at the same density in the large-scale voids far
away from sources. Once heated, the gas cools through adiabatic expansion and Compton scattering. There is only minor additional heat input from subsequent photo-ionization of the small residual neutral fraction because
the recombination time is longer than the Hubble time at low densities. In Fig.~\ref{fig:T_of_z}, we plot the temperature of gas cells with densities $\delta=-0.5\pm0.05$, taken at the end of reionization, as a function of $\zreion$. Indeed, we find a strong correlation whereby low-density gas with earlier $\zreion$ has lower temperatures. Thus, the variations in the temperature at a fixed density in Fig.~\ref{fig:T_of_d} are simply correlated with the timing of reionization. We expect a more negative slope $\gamma-1$ and a larger spread in temperature for more a extended reionization epoch.

Fig.~\ref{fig:T_of_d} also shows the evolved temperature-density relations at $z=4$. For the early model, the scale factor has roughly doubled since the end of reionization and enough time has passed for the photo-ionized IGM to cool and settle into a monotonically rising curve. The $T$-$\rho$ relation for underdense gas can be approximated by a
power-law with an effective slope $\gamma-1\sim0.5$, which is close to the corresponding value calculated by \citet[][see Fig.~4]{1997HuiGnedin} for uniform reionization and smaller than the maximum value of 0.62 obtained when far after reionization. The adiabatic and Compton cooling processes have also significantly reduced the scatter such that the 96\% spread is now less than a thousand degrees K, also similar to what is found for uniform reionization.

However, the $T$-$\rho$ relation for the late model is still relatively complex, with the lowest density gas having an isothermal equation-of-state while near mean density, the effective slope is $\gamma-1\sim0.47$. Furthermore, the scatter is still significant with a 96\% spread of several thousand degrees K. These characteristics are remnants of the complicated conditions at the end of reionization. For uniform reionization, the scatter is several times smaller, approximately a thousand degrees K, and the curvature is also smaller, but still persistent since not enough time has passed for cooling to set up the power-law relation \citep[e.g.][]{1997HuiGnedin, 2004Trac}. It is the scatter rather than the curvature where we find the dominant difference because at the end of reionization, the inversion of the $T$-$\rho$ relation is weak and not far from isothermal. 

Interestingly, we find that the $T$-$\rho$ relation for the early model at $z=6$ (not shown in figures) is similar to that of the late model at $z=4$. Thus, even for early reionization models that end at $z\sim9$, the inhomogeneity of hydrogen reionization may still be imprinted in the spectra of high redshift quasars. Since the scatter in temperature is not random, but correlated with the redshift of reionization, which in turn is correlated with the large-scale density field, it may be possible to probe the correlated inhomogeneity with the $\lya$ forest through statistics such as the flux power spectrum \citep[e.g.][]{2004FangWhite}.

\section{Conclusions}

We have studied the photo-ionization and photo-heating of the IGM from hydrogen reionization using two hydrodynamic + RT simulations. We considered two basic models in which reionization is completed early ($z\sim9$) and late ($z\sim6$) and found for both models that the temperature of a low-density region at a later time is inversely related to the redshift of reionization of that region. low-density gas near large-scale overdensities is ionized and heated earlier than gas in the large-scale, underdense voids. As a result, at the end of reionization the median temperature-density relation is an inverted power-law with slope $\gamma-1\sim-0.2$. There is up to order unity scatter in the temperature at fixed density due to the distribution of reionization redshifts. We expect that both the slope and scatter will depend on the reionization history, especially its duration. Furthermore, it is known that radiative transfer effects from Lyman limit systems can modify the temperature distribution \citep[e.g.][]{1999AbelHaehnelt}, although more for helium rather than hydrogen reionization because for the latter, the UV spectrum from stellar sources is relatively soft and only weak spectral filtering can occur.

We conclude that at the high redshift range $4\la z\la 6$, the equation-of-state of the IGM is more complicated than the commonly assumed forms (isothermal or a tight power-law relation). It is important to keep this result in mind when interpreting the Ly$\alpha$ absorption in quasar spectra \citep[e.g.][]{2002Fan, 2006Fan, 2006Lidz, 2008Gallerani}. In fact, a recent analysis by \citet{2007Becker} has already suggested that adopting an inverted temperature-density relation instead of an isothermal model may have profound implications on the interpretation of the reionization process based on the high redshift Ly$\alpha$ forest. We will study the observational signatures of inhomogeneous hydrogen reionization in an upcoming paper.

\acknowledgments

We thank C.-A.~Faucher-Gigu\`ere, L.~Hernquist, A.~Lidz, M.~McQuinn, and M.~Zaldarriaga for many stimulating discussions. We also thank J.~Chang at NASA for invaluable supercomputing support, N.~Gnedin for his compilation of ionization and recombination rates, and D.~Schaerer for the Pop II SEDs. This work is supported in part by NASA grants NNX08AL43G and NNG06GI09G, NSF grant AST-0407176, FQXi, and Harvard University funds. Computing resources were in part provided by the NASA High-End Computing (HEC) Program through the NASA Advanced Supercomputing (NAS) Division at Ames Research Center.

\bibliographystyle{apj}
\bibliography{astro}

\begin{thebibliography}{44}
\expandafter\ifx\csname natexlab\endcsname\relax\def\natexlab#1{#1}\fi

\bibitem[{{Abel} \& {Haehnelt}(1999)}]{1999AbelHaehnelt}
{Abel}, T., \& {Haehnelt}, M.~G. 1999, \apjl, 520, L13

\bibitem[{{Barkana} \& {Loeb}(2004)}]{2004BarkanaLoeb}
{Barkana}, R., \& {Loeb}, A. 2004, \apj, 609, 474

\bibitem[{{Becker} {et~al.}(2007){Becker}, {Rauch}, \& {Sargent}}]{2007Becker}
{Becker}, G.~D., {Rauch}, M., \& {Sargent}, W.~L.~W. 2007, \apj, 662, 72

\bibitem[{{Bolton} {et~al.}(2008){Bolton}, {Oh}, \& {Furlanetto}}]{2008bBolton}
{Bolton}, J.~S., {Oh}, S.~P., \& {Furlanetto}, S.~R. 2008, ArXiv e-prints, 807

\bibitem[{{Cen}(2003)}]{2003cenb}
{Cen}, R. 2003, \apj, 591, 12

\bibitem[{{Cohn} \& {White}(2008)}]{2008CohnWhite}
{Cohn}, J.~D., \& {White}, M. 2008, \mnras, 385, 2025

\bibitem[{{Dunkley} {et~al.}(2008){Dunkley}, {Komatsu}, {Nolta}, {Spergel},
  {Larson}, {Hinshaw}, {Page}, {Bennett}, {Gold}, {Jarosik}, {Weiland},
  {Halpern}, {Hill}, {Kogut}, {Limon}, {Meyer}, {Tucker}, {Wollack}, \&
  {Wright}}]{2008Dunkley}
{Dunkley}, J., {Komatsu}, E., {Nolta}, M.~R., {Spergel}, D.~N., {Larson}, D.,
  {Hinshaw}, G., {Page}, L., {Bennett}, C.~L., {Gold}, B., {Jarosik}, N.,
  {Weiland}, J.~L., {Halpern}, M., {Hill}, R.~S., {Kogut}, A., {Limon}, M.,
  {Meyer}, S.~S., {Tucker}, G.~S., {Wollack}, E., \& {Wright}, E.~L. 2008,
  ArXiv e-prints, 803

\bibitem[{{Fan} {et~al.}(2006{\natexlab{a}}){Fan}, {Carilli}, \&
  {Keating}}]{2006FCK}
{Fan}, X., {Carilli}, C.~L., \& {Keating}, B. 2006{\natexlab{a}}, \araa, 44,
  415

\bibitem[{{Fan} {et~al.}(2002){Fan}, {Narayanan}, {Strauss}, {White}, {Becker},
  {Pentericci}, \& {Rix}}]{2002Fan}
{Fan}, X., {Narayanan}, V.~K., {Strauss}, M.~A., {White}, R.~L., {Becker},
  R.~H., {Pentericci}, L., \& {Rix}, H.-W. 2002, \aj, 123, 1247

\bibitem[{{Fan} {et~al.}(2006{\natexlab{b}}){Fan}, {Strauss}, {Becker},
  {White}, {Gunn}, {Knapp}, {Richards}, {Schneider}, {Brinkmann}, \&
  {Fukugita}}]{2006Fan}
{Fan}, X., {Strauss}, M.~A., {Becker}, R.~H., {White}, R.~L., {Gunn}, J.~E.,
  {Knapp}, G.~R., {Richards}, G.~T., {Schneider}, D.~P., {Brinkmann}, J., \&
  {Fukugita}, M. 2006{\natexlab{b}}, \aj, 132, 117

\bibitem[{{Fang} \& {White}(2004)}]{2004FangWhite}
{Fang}, T., \& {White}, M. 2004, \apjl, 606, L9

\bibitem[{{Furlanetto} \& {Oh}(2008)}]{2008Furlanetto}
{Furlanetto}, S.~R., \& {Oh}, S.~P. 2008, \apj, 681, 1

\bibitem[{{Furlanetto} {et~al.}(2004){Furlanetto}, {Zaldarriaga}, \&
  {Hernquist}}]{2004FZH}
{Furlanetto}, S.~R., {Zaldarriaga}, M., \& {Hernquist}, L. 2004, \apj, 613, 1

\bibitem[{{Gallerani} {et~al.}(2008){Gallerani}, {Ferrara}, {Fan}, \&
  {Choudhury}}]{2008Gallerani}
{Gallerani}, S., {Ferrara}, A., {Fan}, X., \& {Choudhury}, T.~R. 2008, \mnras,
  386, 359

\bibitem[{{Gnedin}(2000)}]{2000bGnedin}
{Gnedin}, N.~Y. 2000, \apj, 542, 535

\bibitem[{{Gnedin} \& {Hui}(1998)}]{1998GnedinHui}
{Gnedin}, N.~Y., \& {Hui}, L. 1998, \mnras, 296, 44

\bibitem[{{Haiman} \& {Holder}(2003)}]{2003HaimanHolder}
{Haiman}, Z., \& {Holder}, G.~P. 2003, \apj, 595, 1

\bibitem[{{Heitmann} {et~al.}(2007){Heitmann}, {Lukic}, {Fasel}, {Habib},
  {Warren}, {White}, {Ahrens}, {Ankeny}, {Armstrong}, {O'Shea}, {Ricker},
  {Springel}, {Stadel}, \& {Trac}}]{2007Heitmann}
{Heitmann}, K., {Lukic}, Z., {Fasel}, P., {Habib}, S., {Warren}, M.~S.,
  {White}, M., {Ahrens}, J., {Ankeny}, L., {Armstrong}, R., {O'Shea}, B.,
  {Ricker}, P.~M., {Springel}, V., {Stadel}, J., \& {Trac}, H. 2007, ArXiv
  e-prints, 706

\bibitem[{{Hui} \& {Gnedin}(1997)}]{1997HuiGnedin}
{Hui}, L., \& {Gnedin}, N.~Y. 1997, \mnras, 292, 27

\bibitem[{{Hui} \& {Haiman}(2003)}]{2003HuiHaiman}
{Hui}, L., \& {Haiman}, Z. 2003, \apj, 596, 9

\bibitem[{{Lee} {et~al.}(2008){Lee}, {Cen}, {Gott}, \& {Trac}}]{2008Lee}
{Lee}, K.-G., {Cen}, R., {Gott}, J.~R.~I., \& {Trac}, H. 2008, \apj, 675, 8

\bibitem[{{Lewis} {et~al.}(2000){Lewis}, {Challinor}, \& {Lasenby}}]{2000Lewis}
{Lewis}, A., {Challinor}, A., \& {Lasenby}, A. 2000, \apj, 538, 473

\bibitem[{{Lidz} {et~al.}(2006){Lidz}, {Oh}, \& {Furlanetto}}]{2006Lidz}
{Lidz}, A., {Oh}, S.~P., \& {Furlanetto}, S.~R. 2006, \apjl, 639, L47

\bibitem[{{Loeb}(2008)}]{2008Loeb}
{Loeb}, A. 2008, ArXiv e-prints, 804

\bibitem[{{Luki{\'c}} {et~al.}(2007){Luki{\'c}}, {Heitmann}, {Habib},
  {Bashinsky}, \& {Ricker}}]{2007LukicHHBR}
{Luki{\'c}}, Z., {Heitmann}, K., {Habib}, S., {Bashinsky}, S., \& {Ricker},
  P.~M. 2007, \apj, 671, 1160

\bibitem[{{McDonald} {et~al.}(2001){McDonald}, {Miralda-Escud{\' e}}, {Rauch},
  {Sargent}, {Barlow}, \& {Cen}}]{2001McDonald}
{McDonald}, P., {Miralda-Escud{\' e}}, J., {Rauch}, M., {Sargent}, W.~L.~W.,
  {Barlow}, T.~A., \& {Cen}, R. 2001, \apj, 562, 52

\bibitem[{{McQuinn} {et~al.}(2008){McQuinn}, {Lidz}, {Zaldarriaga},
  {Hernquist}, {Hopkins}, {Dutta}, \& {Faucher-Giguere}}]{2008McQuinn}
{McQuinn}, M., {Lidz}, A., {Zaldarriaga}, M., {Hernquist}, L., {Hopkins},
  P.~F., {Dutta}, S., \& {Faucher-Giguere}, C.~. 2008, ArXiv e-prints, 807

\bibitem[{{Miralda-Escude} \& {Ostriker}(1990)}]{1990MiraldaEscude}
{Miralda-Escude}, J., \& {Ostriker}, J.~P. 1990, \apj, 350, 1

\bibitem[{{Miralda-Escud{\'e}} \& {Rees}(1994)}]{1994MiraldaEscude}
{Miralda-Escud{\'e}}, J., \& {Rees}, M.~J. 1994, \mnras, 266, 343

\bibitem[{{Reed} {et~al.}(2007){Reed}, {Bower}, {Frenk}, {Jenkins}, \&
  {Theuns}}]{2007ReedBFJT}
{Reed}, D.~S., {Bower}, R., {Frenk}, C.~S., {Jenkins}, A., \& {Theuns}, T.
  2007, \mnras, 374, 2

\bibitem[{{Ricotti} {et~al.}(2000){Ricotti}, {Gnedin}, \&
  {Shull}}]{2000Ricotti}
{Ricotti}, M., {Gnedin}, N.~Y., \& {Shull}, J.~M. 2000, \apj, 534, 41

\bibitem[{{Schaerer}(2003)}]{2003Schaerer}
{Schaerer}, D. 2003, \aap, 397, 527

\bibitem[{{Schaye} {et~al.}(1999){Schaye}, {Theuns}, {Leonard}, \&
  {Efstathiou}}]{1999Schaye}
{Schaye}, J., {Theuns}, T., {Leonard}, A., \& {Efstathiou}, G. 1999, \mnras,
  310, 57

\bibitem[{{Schaye} {et~al.}(2000){Schaye}, {Theuns}, {Rauch}, {Efstathiou}, \&
  {Sargent}}]{2000Schaye}
{Schaye}, J., {Theuns}, T., {Rauch}, M., {Efstathiou}, G., \& {Sargent},
  W.~L.~W. 2000, \mnras, 318, 817

\bibitem[{{Theuns} {et~al.}(2000){Theuns}, {Schaye}, \&
  {Haehnelt}}]{2000Theuns}
{Theuns}, T., {Schaye}, J., \& {Haehnelt}, M.~G. 2000, \mnras, 315, 600

\bibitem[{{Theuns} {et~al.}(2002){Theuns}, {Schaye}, {Zaroubi}, {Kim},
  {Tzanavaris}, \& {Carswell}}]{2002Theuns}
{Theuns}, T., {Schaye}, J., {Zaroubi}, S., {Kim}, T.-S., {Tzanavaris}, P., \&
  {Carswell}, B. 2002, \apjl, 567, L103

\bibitem[{{Trac}(2004)}]{2004Trac}
{Trac}, H. 2004, PhD thesis, University of Toronto (Canada), Canada

\bibitem[{{Trac} \& {Cen}(2007)}]{2007TracCen}
{Trac}, H., \& {Cen}, R. 2007, \apj, 671, 1

\bibitem[{{Trac} \& {Pen}(2004)}]{2004TracPen}
{Trac}, H., \& {Pen}, U.-L. 2004, New Astronomy, 9, 443

\bibitem[{{Trac} \& {Pen}(2006)}]{2006TracPen}
---. 2006, New Astronomy, 11, 273

\bibitem[{{Viel} {et~al.}(2004){Viel}, {Haehnelt}, \& {Springel}}]{2004Viel}
{Viel}, M., {Haehnelt}, M.~G., \& {Springel}, V. 2004, \mnras, 354, 684

\bibitem[{{Wyithe} \& {Loeb}(2003)}]{2003WyitheLoeb}
{Wyithe}, J.~S.~B., \& {Loeb}, A. 2003, \apj, 586, 693

\bibitem[{{Wyithe} \& {Loeb}(2004)}]{2004WyitheLoeb}
---. 2004, \nat, 432, 194

\bibitem[{{Zaldarriaga} {et~al.}(2001){Zaldarriaga}, {Hui}, \&
  {Tegmark}}]{2001Zaldarriaga}
{Zaldarriaga}, M., {Hui}, L., \& {Tegmark}, M. 2001, \apj, 557, 519

\end{thebibliography}

\end{document}